\documentclass[journal,romanappendices]{IEEEtran}

\usepackage{amsmath,amsfonts,amsthm,amssymb}
\usepackage{booktabs}
\usepackage{cite}

\newcommand{\bx}{\mathbf{x}}
\newcommand{\by}{\mathbf{y}}

\newtheorem{theorem}{Theorem}
\newtheorem{lemma}{Lemma}

\newtheorem{proposition}{Proposition}%
\newtheorem{problem}{Problem}

\newcommand{\avg}{\Gamma}
\newcommand{\transpose}{\mathrm{T}}
\newcommand{\hb}{h_{\mathrm{b}}}

\begin{document}

\title{Tight Bound on Randomness for Violating the
  Clauser-Horne-Shimony-Holt Inequality} %

\author{Yifeng~Teng,~Shenghao~Yang,~\IEEEmembership{Member,~IEEE},~Siwei~Wang~and~Mingfei~Zhao%
\thanks{This work was
  supported in part by the National Natural Science Foundation of
  China (NSFC) under Grant 61471215. This work was partially funded by
  a grant from the University Grants Committee of the Hong Kong
  Special Administrative Region (Project No.\ AoE/E-02/08).}
\thanks{Y. Teng is with the Department of Computer Sciences,
  University of Wisconsin-Madison, Madison, USA (e-mail:
  yifengt@cs.wisc.edu).}
\thanks{S. Yang is with the School of Science of Engineering, The
  Chinese University of Hong Kong, Shenzhen, P. R. China (e-mail:
  shyang@cuhk.edu.cn).}
\thanks{S. Wang is with the Institute for Interdisciplinary
  Information Sciences, Tsinghua University, Beijing, P. R. China
  (e-mail: wangsw15@mails.tsinghua.edu.cn).}
\thanks{M. Zhao is with the School of Computer Science, McGill
  University, Montreal, Canada (e-mail:
  mingfei.zhao@mail.mcgill.ca).}%
}

\maketitle

\begin{abstract}
  Free will (or randomness) has been studied to achieve loophole-free Bell's
  inequality test and to provide device-independent quantum key
  distribution security proofs. The required randomness
  such that a local hidden variable model (LHVM) can violate the
  Clauser-Horne-Shimony-Holt (CHSH) inequality has been studied, but a
  tight bound has not been proved for a practical case that i) the
  device settings of the two parties in the Bell test are independent;
  and ii) the device settings of each party can be
  correlated or biased across different runs. Using some information theoretic techniques, we prove in this paper a tight bound on the required
  randomness for this case such that the CHSH inequality can be
  violated by certain LHVM. %
  Our proof has a clear achievability and converse style. The achievability part is proved using type counting. To prove the converse part, we introduce a concept called profile for a set of binary sequences and study the properties of profiles. Our profile-based converse technique is also of independent interest.
\end{abstract}


\section{Introduction}

Bell's inequality test \cite{bell1964einstein} provides an approach to
verify the existence of physical phenomenon that cannot be explained by
local hidden variable models (LHVMs).  The Clauser-Horne-Shimony-Holt
(CHSH) inequality \cite{Clauser69} is the most often used inequality
in Bell test experiments.  Experimental demonstrations of the
violation of CHSH inequalities have been conducted since 1982
\cite{Aspect82} (see also Giustina et al.'s work \cite{Giustina13} and
the references therein).  These Bell tests, however, suffer from an
inherent loophole that the settings of the participated devices may
not be chosen totally randomly, called the \emph{randomness (free
  will)} loophole.  A small amount of correction between the device
settings makes it possible that a LHVM
can reproduce predictions of quantum mechanics
\cite{Feldmann95,Kofler06,Hall10,Barrett11,Hall11}.
This loophole also weakens the Bell's inequality based security proofs
of device-independent quantum key distribution
\cite{Mayers1998,Acin06,Vazirani14} and randomness expansion
\cite{Pironio10,Colbeck12,Dhara14}.

One of the essential questions in the randomness loophole is the bound
of randomness such that the correctness of Bell tests can (or cannot)
be guaranteed
\cite{Hall10,Barrett11,Hall11,Koh12,Thinh13,Pop13,yuan2014randomness,Putz14}. Using
a min-entropy type randomness measure, the bound of randomness
required in a CHSH inequality test can be formulated as an
optimization problem, and various special cases have been solved
\cite{Koh12,Pop13,yuan2014randomness}. One case that has not been
completely resolved in the literature is that the two parties of the
test have independent settings, but the setting of each party can be
biased or correlated across different runs. In this paper, we study
this case and obtain the asymptotic optimal value explicitly.

\subsection{Problem Formulation}

Let $n$ be a positive integer, and $X, Y$ be two random variables over
$\{0,1\}^n$ with a joint distribution $p_{XY}$.  We may consider that
$X$ and $Y$ are the device settings of the two parties in an $n$-run
Bell test, respectively.  The following randomness measure has been used in the literature:
\begin{equation*}
  P = \left(\max_{\bx,\by\in\{0,1\}^n}p_{XY}(\bx,\by) \right)^{1/n}.
\end{equation*}
When $X$ and $Y$ are independent and uniformly distributed, $P =
1/4$, which is the minimum value of $P$ and corresponds to the
case of complete randomness. When $X$ and $Y$ are deterministic,
$P=1$, which corresponds to the case of zero randomness. Note
that $P$ is related to the min-entropy:
\begin{equation*}
  H_{\infty}(X,Y) := -\log
\max_{\bx,\by\in\{0,1\}^n}p_{XY}(\bx,\by) = -n \log P.
\end{equation*}

Regard the vectors $\bx \in \{0,1\}^n$ as column vectors and denote by
$\bx^\transpose$ the transpose of the $\bx$. The
optimization problem of interest is
\begin{equation} \label{eq:13}
  \begin{array}{cl}
   \displaystyle{\min_{p_{XY}}} & P \\
  \text{s.t.} & \displaystyle\frac{1}{n} \sum_{\bx,\by}\bx^{\transpose}\by p_{XY}(\bx,\by)\leq \frac{4-S_Q}{8},
  \end{array}
\end{equation}
where $S_Q=2\sqrt{2}$ is a quantum constant. Readers may refer to
\cite{Hall10,Koh12,Pop13} to see how this problem is obtained. Optimization
\eqref{eq:13} can be simplified to a linear programming \cite{Pop13}. 
When $n=1$, the optimal value of \eqref{eq:13} is
$(S_Q + 4)/24 \approx 0.285$, which was shown by Hall \cite{Hall10} and Koh
et al. \cite{Koh12}. When $n\rightarrow \infty$, Pope and Kay
\cite{Pop13} showed that the optimal value of \eqref{eq:13} converges
to $3^{\frac{-S_Q-4}{8}}2^{-\hb\left(\frac{4-S_Q}{8}\right)} \approx
0.258$, where
\begin{equation*}
  \hb(t)=-t\log_2t-(1-t)\log_2(1-t)
\end{equation*}
is the binary entropy function. %

The case that $X$ and $Y$ are independent is of particular
interest. Towards a loophole free Bell test, physicists have designed
experiments with independent device settings \cite{Gallicchio14}.  In
quantum key distribution, the experimental devices of the two parties
may be manufactured independently and separated spatially, reducing
the potential correlation of the device settings generated by the
adversary. For independent device settings, the corresponding
optimization problem becomes
\begin{equation} \label{eq:7}
  \begin{array}{cl}
   \displaystyle{\min_{p_{X},p_{Y}}} & P \\
  \text{s.t.} & \displaystyle\frac{1}{n}
                \sum_{\bx,\by}\bx^{\transpose}\by
                p_{XY}(\bx,\by)\leq \frac{4-S_Q}{8}\\
    & p_{XY}(\bx,\by)  = p_{X}(\bx)p_{Y}(\by).
  \end{array}
\end{equation}
Note that the above problem is not derived by directly imposing the
constraint $p_{XY}(\bx,\by) = p_{X}(\bx)p_{Y}(\by)$ to \eqref{eq:13}.
For the completeness, we briefly discuss how \eqref{eq:7} is derived
from the corresponding CHSH inequality test problem in
Appendix~\ref{sec:deriv-optim-probl}.

When $n=1$, it was obtained by Koh et al. \cite{Koh12} that the
optimal value of \eqref{eq:7} is $S_Q/8 \approx 0.354$. %
Let $P_Q$ be the limit of the optimal value of \eqref{eq:7}, when
$n\rightarrow \infty$. The value of $P_Q$ has the following
interpretation. For any independent device settings with randomness
less than $P_Q$, it is not possible to have a LHVM that violates CHSH
inequality. But for any value $P>P_Q$, there exists a LHVM that
violates CHSH inequality where the device settings are independent,
but have randomness less than or equal to $P$. Therefore, we are
motivated to study the value of $P_Q$ for CHSH inequality test.  Yuan,
Cao and Ma \cite{yuan2014randomness} have shown numerically that
$P_Q \lessapprox 0.264$.

\subsection{Our Contribution}

In this paper, we provide an exact characterization of $P_Q$, and
hence  close the unresolved case in Table~\ref{tab:1}. Particularly, we
show that 
\begin{equation*}
  P_Q = 4^{-\hb(\sqrt{c_Q})} = 0.26428\ldots,
\end{equation*}
where \(c_Q=\frac{4-S_Q}{8} \approx 0.1464\).
Our formula has a min-entropy interpretation:
$-n\log_2 P_Q = 2n \hb(\sqrt{c_Q})$, i.e., each bit in $X$ and $Y$ has
an average min-entropy $\hb(\sqrt{c_Q})$. 

To prove achievability, we simplify \eqref{eq:7} by introducing an extra
constraint that both $X$ and $Y$ have the uniform distribution over
$\mathcal{A}_{n,l}$, the set of sequences in $\{0,1\}^n$ with at most $nl$ $1$s,
and obtain a new optimization problem 
\begin{equation} \tag{$2'$} \label{eq:72}
  \begin{array}{cl}
   \displaystyle{\min_{l}} & (1/|\mathcal{A}_{n,l}|)^{2/n} \\
  \text{s.t.} & \displaystyle\frac{1}{n |\mathcal{A}_{n,l}|^2}
                \sum_{\bx,\by\in \mathcal{A}_{n,l}}\bx^{\transpose}\by
                \leq \frac{4-S_Q}{8},
  \end{array}
\end{equation}
which is essentially the same problem studied in
\cite[Section IV-B]{yuan2014randomness}. The asymptotic optimal value
of \eqref{eq:72} when $n \rightarrow \infty$, denoted by $\hat P_Q$, gives an upper bound on $P_Q$
since \eqref{eq:72}
is obtained by reducing the feasible region of \eqref{eq:7}. The
numerical bound on $\hat P_Q$  in \cite{yuan2014randomness} can be made analytical, and it shows that $\hat P_Q\leq 4^{-\hb(\sqrt{c_Q})}$ and hence
$P_Q \leq 4^{-\hb(\sqrt{c_Q})}$.

The major part of our paper is to show the converse that no
distributions of $X$ and $Y$ with randomness less than
$4^{-\hb(\sqrt{c_Q})}$ can be feasible for \eqref{eq:7}, i.e., $P_Q
\geq 4^{-\hb(\sqrt{c_Q})}$. Note that we cannot use \eqref{eq:72} as
the starting point to prove the converse since the derivation of
\eqref{eq:72} implies $\hat{P}_Q \geq P_Q$. It is possible to show that
$\hat{P}_Q \geq 4^{-\hb(\sqrt{c_Q})}$, but
not $P_Q \geq 4^{-\hb(\sqrt{c_Q})}$ by studying only \eqref{eq:72}.

To prove converse, we introduce a concept called \emph{profile} to
characterize a set of binary sequences.  We study some properties of
profiles, based on which optimization \eqref{eq:7} is simplified and the converse is proved. The technique of profile seems to be firstly used here and may of independent interest for other problems.

In the remainder of this paper, our techniques used to prove the main result are summarized in the
next section, followed by the details in
Section~\ref{sec:proofs-main-results}. Some concluding remarks are given in Section~\ref{sec:concluding-remarks}.

\begin{table}
\centering
\caption{\label{tab:1} Previous results.}
\begin{tabular}{ccc}
  \hline\hline
  & correlated devices & independent devices \\
  \hline
  $n=1$ & $(S_Q+4)/24 \approx 0.285$ & $S_Q/8 \approx 0.354$ \\
  $n\rightarrow\infty$ &
  $3^{-\frac{S_Q+4}{8}}2^{-\hb\left(\frac{4-S_Q}{8}\right)} \approx 0.258$
  & $ \lessapprox 0.264$ \\
  \hline\hline
\end{tabular}
\end{table}

\section{Outline of the Proofs}
As described in the previous section, we formulate an optimization problem as follows.

\begin{problem}\label{problem1}
For any given $c\in (0,1/4]$ and every positive integer $n$, consider the following program
\begin{equation*}
\begin{array}{cl}
  \displaystyle{\min_{p_X,p_Y}} & \displaystyle{\left(\max_{\bx}p_X(\bx)
      \max_{\by}p_Y(\by)\right)^{1/n}},\nonumber \\
  \text{s.t.} & \displaystyle{\frac{1}{n}\sum_{\bx,\by\in \{0,1\}^n}p_X(\bx)p_Y(\by)\bx^\transpose\by\leq c},
\end{array}
\end{equation*}
where $p_X$ and $p_Y$ are probability distributions over $\{0,1\}^n$.
Let $P_n$ be the optimal
value of the above program.  We are interested in the limit of the
sequence $\{P_n\}$ when $n\rightarrow\infty$.
\end{problem}

Specifically we
will need the case that $c=c_Q$ for the physics problem of
interests. Now we state the following theorem.
\begin{theorem}\label{thm1}
For Problem~\ref{problem1} with $c=c_Q$,
\(\displaystyle \lim_{n\to\infty}P_n=4^{-\hb(\sqrt{c_Q})}\), where
\begin{equation}\label{eq:8}
  \hb(t)=-t\log_2t-(1-t)\log_2(1-t)
\end{equation}
is the binary entropy function.
\end{theorem}
In the following of this section, we give an outline of the main
techniques towards proving this theorem.  We have the following bound
for $P_n$.

\begin{proposition}\label{prop:1}
  For all sufficiently large $n$,
  $1/4 \leq  P_n  < 1/2$.
\end{proposition}

\subsection{Simplified Problem}

Let $S_X$ and $S_Y$
be the support of distributions $p_X$ and $p_Y$, respectively.
Problem \ref{problem1} can be simplified if we only consider
distributions that are uniform over support. Suppose that
\begin{eqnarray*}
  p_X(\bx) & = & \frac{1}{|S_X|},\ \forall \bx\in S_X, \\
  p_Y(\by) & = &  \frac{1}{|S_Y|},\ \forall \by\in S_Y.
\end{eqnarray*}
Then we have
\begin{equation*}
  \left(\max_{\bx}p_X(\bx) \max_{\by}p_Y(\by)\right)^{1/n} = \frac{1}{\sqrt[n]{|S_X|\cdot|S_Y|}},
\end{equation*}
and
\begin{equation*}
  \frac{1}{n}\sum_{\bx,\by\in \{0,1\}^n}p_X(\bx)p_Y(\by)\bx^\transpose\by =
  \frac{\sum_{\bx\in S_X,\by\in S_Y}\bx^\transpose\by}{n|S_X|\cdot|S_Y|}.
\end{equation*}
Define a new problem as follows:

\begin{problem}\label{problem2}
For any given $c\in (0,1/4]$ and every positive integer \(n\), consider the following programming
\begin{eqnarray}
  \displaystyle{\min_{S_X,S_Y}} & \  & \displaystyle{\frac{1}{\sqrt[n]{|S_X|\cdot|S_Y|}}},\nonumber \\
  \text{s.t.} & & \displaystyle{\frac{1}{n|S_X|\cdot|S_Y|} \sum_{\bx\in
      S_X,\by\in S_Y}\bx^\transpose\by \leq c}, \label{constraint2}
\end{eqnarray}
where $S_X$ and $S_Y$ are subsets of \(\{0,1\}^n\).
Let $P_n'$ be the optimal value of the above program.
We are interested in the limit of the sequence $\{P_n'\}$ when
$n\rightarrow\infty$.
\end{problem}

It is obvious that $P_n \leq P_n'$ since only distributions that are
uniform over support are considered in Problem~\ref{problem2}. The
following theorem enables us to focus on $\lim_n P_n'$.

\begin{theorem}\label{t2}
$\lim_{n\to\infty}P_n'/P_n=1$.
\end{theorem}

\subsection{Profiles}

To study the properties of a set of binary vectors, we introduce the
concept of \textit{profile}. For any positive integer \(m\), we call
vector \(a=(a_1,a_2,\cdots,a_m)\in[0,1]^m\) a \emph{profile} or an
\textit{m-profile}.
For each \(S\subseteq\{0,1\}^n\), define the \textit{profile} of set \(S\) as
\begin{equation*}
\avg(S)=\begin{cases}
\displaystyle \frac{1}{|S|}\sum_{s\in S}{s},& \text{\(|S|>0\)};\\
(0,0,\ldots,0), & \text{\(|S|=0\)}.
\end{cases}
\end{equation*}
We see that \(\avg(S)\) is an \(n\)-profile.

Define the \emph{characteristic function} of an
$m$-profile \(a\) as \(\displaystyle f_a:[0,1]\to[0,1]\) such that
\begin{equation*}
f_a(t)=\begin{cases}
a_1,& t=0;\\
a_{\lceil tm\rceil},& \forall 0<t\leq 1.
\end{cases}
\end{equation*}
The characteristic function of a profile is a step function. 
For two profiles \(a\) and \(b\), we say \(a\leq b\) if for any
\(0\leq r\leq 1\), \(f_a(r)\leq f_b(r)\), where $a$ and $b$ may not
include the same number of components. 
For a vector $a$, we denote by $a_i$ the $i$-th component of $a$.
\begin{lemma}\label{lm:1}
  For two $n$ profiles $a$ and $b$, $\frac{1}{n} a^{\transpose}b =\int_{0}^{1}{f_{a}(t)f_{b}(t)}\mathrm{d}t$.
\end{lemma}
\begin{IEEEproof}
We write according to the definition that
  \begin{IEEEeqnarray*}{rCl}
\frac{1}{n}a^{\transpose}b 
&=&\frac{1}{n} \sum_{i=1}^n a_ib_i \\
&=&\frac{1}{n} \sum_{i=1}^n \int_{(i-1)/n}^{i/n} f_a(t) f_b(t) \mathrm{d}t \\
&=&\int_{0}^{1}{f_{a}(t)f_{b}(t)}\mathrm{d}t,
  \end{IEEEeqnarray*}
where the second equality holds due to the fact that the characteristic
function of a profile is a step function. 
\end{IEEEproof}

The following lemma tells us how to represent the
constraint in Problem~\ref{problem2} in a simple way using profiles.

\begin{lemma} \label{lemma4}
In Problem~\ref{problem2}, the left hand side of constraint (\ref{constraint2}) can be expressed as
\begin{equation*}
  \frac{1}{n|S_X|\cdot|S_Y|} \sum_{\bx\in S_X,\by\in S_Y}\bx^\transpose\by =
  \frac{1}{n} a^{\transpose}b,
\end{equation*}
where \(a=\avg(S_X)\) and \(b=\avg(S_Y)\).
\end{lemma}
\begin{IEEEproof}
We can write
\begin{IEEEeqnarray*}{rCl}
\IEEEeqnarraymulticol{3}{l}{\frac{1}{n}\cdot\frac{1}{|S_X|}\cdot\frac{1}{|S_Y|}\sum_{\bx\in
    S_X, \by\in S_Y}\bx^\transpose\by} \\ \quad
&=& \frac{1}{n}\cdot\frac{1}{|S_X|}\cdot\frac{1}{|S_Y|}\left(\sum_{\bx\in
    S_X}{\bx}\right)^\transpose\left(\sum_{\by\in S_Y}{\by}\right) 
\\
&=&\frac{1}{n}\cdot\frac{1}{|S_X|}\cdot\frac{1}{|S_Y|}(|S_X|a)^{\transpose}(|S_Y|b) \IEEEyesnumber
\label{eq:ds} \\
&=&\frac{1}{n}a^{\transpose}b, 
\end{IEEEeqnarray*}
where \eqref{eq:ds} follows from the definition of the profile
of a set of binary vectors.
\end{IEEEproof}

The following theorem states that to get the value of \(P_n'\), we
only need to consider \(S_X\) and \(S_Y\) with certain monotone
property of their profiles.
\begin{theorem}\label{t3}
For all n, there exist $S_X, S_Y\subseteq \{0,1\}^n$ that achieve
\(P_n'\) in Problem \ref{problem2} such that for \(a=\avg(S_X)\) and
\(b=\avg(S_Y)\), \(0.5\geq a_1\geq a_2\geq...\geq a_n\geq0\)  and \(0\leq b_1\leq b_2\leq...\leq b_n\leq 0.5\).
\end{theorem}

By Theorem \ref{t3}, it is sufficient for us to consider only profiles \(a\in[0,0.5]^m\).
For each \(m\)-profile \(a\), define its \textit{$n$-volume} to be
\begin{equation}
V_n(a)=\max\left\{|S|:S\subseteq\{0,1\}^n,\ \avg(S)\leq a\right\},
\end{equation}
where $n$ may not be the same as $m$.

\begin{lemma}\label{volumelemma}
For any two profiles \(p\) and \(q\), if \(p\leq q\), we have \(V_n(p)\leq V_n(q)\) for every positive integer \(n\).
\end{lemma}
\begin{IEEEproof}
Notice that for any \(n\), any \textit{n-profile} smaller than \(p\) is smaller than \(q\), then the lemma suffices.
\end{IEEEproof}

The following theorem gives an upper bound on the volume of a profile,
which will be used in the proof of the lower bound on $P_n'$.

\begin{theorem}\label{theorem1}
Fix an integer $m$ and
let \(\displaystyle a\in\left[0,0.5\right]^m\) be an
\(m\)-profile. For any positive integer \(n\), the \textit{n-volume} of profile \(a\) satisfies
\begin{equation}
V_n(a)\leq 2^{\frac{n}{m}(\sum_{i=1}^{m}{\hb(a_i)}+o(1))},
\end{equation}
where $\hb$ is the binary entropy function defined in \eqref{eq:8} and
$o(1)\rightarrow 0$ as $n\rightarrow\infty$.
\end{theorem}

\subsection{Converse and Achievability}

\begin{theorem}\label{maintheorem}
For any sequence of \(S_X,S_Y\subseteq \{0,1\}^n\) such that
\begin{equation*}
\frac{1}{n|S_X|\cdot|S_Y|} \sum_{\bx\in
      S_X,\by\in S_Y} \bx^{\transpose}\by\leq c_Q,
\end{equation*}
we have
\begin{equation*}
\liminf_{n\to\infty}{\frac{1}{\sqrt[n]{|S_X||S_Y|}}}\geq 4^{-\hb\left(\sqrt{c_Q}\right)}.
\end{equation*}
\end{theorem}

We then give a construction of $S_X$ and $S_Y$ to show that the
bound in Theorem \ref{maintheorem} is tight.
\begin{theorem}\label{finaltheorem}
There exists a sequence of \(S_X,S_Y\subseteq \{0,1\}^n\) such that
\begin{equation*}
\frac{1}{n|S_X|\cdot|S_Y|} \sum_{\bx\in
      S_X,\by\in S_Y}\bx^{\transpose}\by\leq c_Q,
\end{equation*}
and
\begin{equation*}
\lim_{n\to\infty}{\frac{1}{\sqrt[n]{|S_X||S_Y|}}}=4^{-\hb\left(\sqrt{c_Q}\right)}.
\end{equation*}
\end{theorem}

Now we are ready to prove Theorem~\ref{thm1}.

\begin{IEEEproof}[Proof of Theorem~\ref{thm1}]
Theorem \ref{maintheorem} implies that
\begin{equation*}
  \liminf_{n\to\infty} P_n' \geq
4^{-\hb\left(\sqrt{c_Q}\right)},
\end{equation*}
 and Theorem \ref{finaltheorem} implies that
 \begin{equation*}
   \limsup_{n\to\infty} P_n' \leq
4^{-\hb\left(\sqrt{c_Q}\right)}.
 \end{equation*}
Thus $\lim_{n\to\infty} P_n' = 4^{-\hb\left(\sqrt{c_Q}\right)}$, which
together with Theorem~\ref{t2} proves Theorem \ref{thm1}.
\end{IEEEproof}

\section{Proofs}
\label{sec:proofs-main-results}

\subsection{Proof of Proposition~\ref{prop:1}}

The lower bound follows from $\max_\bx p_X(\bx) \geq 1/2^n$ for any
distribution $p_X$ over $\{0,1\}^n$. To prove the upper bound,
consider the following two distributions:
  \begin{equation*}
    p_X(\bx) =
    \begin{cases}
      1-2c, & \bx = \mathbf{0} \\
      2c/(2^n-1), & \bx \neq \mathbf{0},
    \end{cases}
  \end{equation*}
  where $c\in (0,1/4]$ as given in Problem~\ref{problem2},
  and $p_Y(\by)=1/2^n$ for all $\by\in\{0,1\}^n$. We the have
  \begin{IEEEeqnarray*}{rCl}
   \IEEEeqnarraymulticol{3}{l}{\frac{1}{n}\sum_{\bx,\by\in
      \{0,1\}^n}p_X(\bx)p_Y(\by)\bx^\transpose\by} \\ \quad & = &
    \frac{2c}{2^n(2^n-1)} \cdot \frac{1}{n}\sum_{\bx,\by\in
      \{0,1\}^n} \bx^\transpose\by \\
    & = &  \frac{c }{2^{n-1}(2^n-1)} 2^{2(n-1)} \leq c,
  \end{IEEEeqnarray*}
and
\begin{IEEEeqnarray*}{rCl}
  \IEEEeqnarraymulticol{3}{l}{P_n \leq \left(\max_{\bx}p_X(\bx)
    \max_{\by}p_Y(\by)\right)^{1/n}} \\ \quad & = & \frac{1}{2} \left(\max\{1-2c,
    2c/(2^n-1)\} \right)^{1/n}\\
  & = & \frac{1}{2}(1-2c)^{1/n} < \frac{1}{2},
\end{IEEEeqnarray*}
where the second equality follows from $c\leq 1/4$ and the last
inequality follows from $c>0$.

\subsection{Proof of Theorem \ref{t2}}

Suppose that $p_X$ and $p_Y$ on $\{0,1\}^n$ achieve the minimum objective value $P_n$ in Problem \ref{problem1}. Write
\begin{equation*}
\sum_{\bx,\by\in
  \{0,1\}^n}p_X(\bx)p_Y(\by)\bx^{\transpose}\by=\sum_\bx p_X(\bx)\theta_{p_Y}(\bx),
\end{equation*}
where
\begin{equation*}
   \theta_{p_Y}(\bx)=\bx^{\transpose}\left(\sum_\by p_Y(\by)\by\right).
\end{equation*}

Let $P_X=\max_\bx p_X(\bx)$. We know that $P_X>0$. If $P_X=1$, then
there exists $\bx_0$ such that $p_X(\bx_0) = 1$. In this case,
$P_n=1/2$ since otherwise we may instead choose $p_X$ such that
$p_X(\mathbf{0})=1$ and $p_Y$ such that $p_Y(\by)=1/2^n$ for all
$\by\in\{0,1\}^n$. Thus we have a contradiction to $P_n<1/2$ (see
Proposition~\ref{prop:1}). Therefore, $0<P_X<1$.

Now consider the following linear program:
\begin{equation}\label{eq:9a}
\begin{aligned}
\min_{p_X} \qquad &\sum_\bx p_X(\bx) \theta_{p_Y}(\bx),\\
\text{s.t.}\qquad &p_X(\bx)\leq P_X, \ \forall \bx\in\{0,1\}^n.
\end{aligned}
\end{equation}
Let $p_X^{*}$ be an optimal distribution that minimizes the objective
of \eqref{eq:9a}. Since the linear program must achieve its optimal value 
at the extreme points, there must be $\lfloor\frac{1}{P_X}\rfloor$
sequences $\bx$ with $p_X^*(\bx)=P_X$ and one
sequence $\mathbf{z}$ with $p_X^*(\mathbf{z}) = 1-\lfloor\frac{1}{P_X}\rfloor
P_X$. For any other sequence $\bx$, we have $p_X^*(\bx)=0$.

We then have
\begin{equation*}
\sum_\bx p_X^{*}(\bx)\theta_{p_Y}(\bx)\leq \sum_\bx p_X(\bx)\theta_{p_Y}(\bx)\leq nc,
\end{equation*}
and
\begin{equation*}
\left(\max_\bx p_X^{*}(\bx)\max_\by p_Y(\by)\right)^{1/n}=\left(P_X\cdot\max_{\by}p_Y(\by)\right)^{1/n}=P_n.
\end{equation*}
Therefore, $p_X^{*}$ and $p_Y$ also obtain the minimum objective value
$P_n$ in Problem~\ref{problem1}.

Let $S_X$ be the support of $p_X^*$. We have $|S_X|= \lceil \frac{1}{P_X}
\rceil$, and for any $\bx \in S_X$, $\theta_{p_Y}(\mathbf{z}) \geq
\theta_{p_Y}(\bx)$.
Let $\bar p_X$ be the uniform distribution over $S_X\backslash\{\mathbf{z}\}$. Notice for all $\bx \in S_X\backslash\{\mathbf{z}\}$,
\begin{equation*}
\bar p_X(\bx) \geq p_X^{*}(\bx),
\end{equation*}
and
\begin{equation*}
\sum_{\bx\in S_X\backslash\{\mathbf{z}\}}(\bar p_X(\bx)-p_X^{*}(\bx))=p_X^{*}(\mathbf{z}).
\end{equation*}
We have
\begin{eqnarray*}
&&\sum_{\bx,\by}\bar p_X(\bx)p_Y(\by)\bx^{\transpose}\by-\sum_{\bx,\by}p_X^{*}(\bx)p_Y(\by)\bx^{\transpose}\by\\
&=&\sum_{\bx\in S_X\backslash\{\mathbf{z}\}}(\bar p_X(\bx)-p_X^{*}(\bx)) \theta_{p_Y}(\bx)
-p_X^{*}(\mathbf{z}) \theta_{p_Y}(\mathbf{z})\\
&\leq &\sum_{x\in S_X\backslash\{\mathbf{z}\}}(\bar p_X(\bx)-p_X^{*}(\bx)) \theta_{p_Y}(\mathbf{z})
-p_X^{*}(\mathbf{z}) \theta_{p_Y}(\mathbf{z})\\
&=& 0.
\end{eqnarray*}
Thus
\begin{equation}\label{eq:10}
  \sum_{\bx,\by}\bar p_X(\bx)p_Y(\by)\bx^{\transpose}\by \leq
  \sum_{\bx,\by}p_X^{*}(\bx)p_Y(\by)\bx^{\transpose}\by \leq nc.
\end{equation}

Let $P_n^\dagger = \min_{p_X,p_Y} \left(\max_{\bx} p_X(\bx)
\max_{\by}p_Y(\by)\right)^{1/n}$ such that $p_X$ and $p_Y$ satisfy the
constraint of Problem~\ref{problem1} and $p_X$ is uniform
over its support. We have
\begin{eqnarray*}
  P_n \leq P_n^\dagger & \leq & \left(\max_{\bx}\bar p_X(\bx)\max_{\by}p_Y(\by)\right)^{1/n}\\
  & = & \left(\frac{1}{\lfloor
      1/P_X\rfloor}\max_{\by}p_Y(\by)\right)^{1/n} \\
  & \leq & \left(\frac{1}{\lceil
      1/P_X\rceil - 1}\max_{\by}p_Y(\by)\right)^{1/n}\\
  & \leq & \left(3P_X\max_{\by}p_Y(\by)\right)^{1/n} \\
  & = & 3^{1/n} P_n,
\end{eqnarray*}
where the second inequality follows from $\bar p_X$ and $p_Y$ satisfy
the constraint of Problem~\ref{problem1} (see \eqref{eq:10}), and the
last inequality follows from $0<P_X<1$ and Lemma~\ref{l3} (to be proved
later in this section).  Therefore, $\lim_{n\rightarrow \infty}
P_n^\dagger/P_n = 1$.

Similar technique can be used to show that $\lim_{n\rightarrow \infty}
P_n'/P_n^\dagger = 1$, which completes the proof of this theorem.
Specifically, suppose that $p_X,p_Y$ on $\{0,1\}^n$ achieve
$P_n^\dagger$ where $p_X$ is uniform on its support.  Define
$P_Y=\max_{\by}p_Y(\by)$ and $P_X=\max_{\bx}p_X(\bx)$. Similar to the
above argument, there exists distribution $p_Y^{*}$ such that
\begin{enumerate}
\item for $\lfloor\frac{1}{P_Y}\rfloor$ sequences $\by$,
  $p_Y^*(\by) = P_Y$, for another one sequence $\by_0$,
  $p_Y^*(\by_0)=1-\lfloor\frac{1}{P_Y}\rfloor P_Y$, and for all other sequences $\by$,
  $p_Y^*(\by)=0$;
\item $\sum_{\bx,\by}p_X(\bx)p_Y^*(\by) \bx^\transpose\by \leq
  \sum_{\bx,\by}p_X(\bx)p_Y(\by) \bx^\transpose\by \leq nc$; and
\item $(\max_\bx p_X(\bx) \max_{\by}p_Y^*(\by))^{1/n} = (P_XP_Y)^{1/n}$.
\end{enumerate}

Let the support set of distributions $p_Y^{*}$ be $S_Y$, and let $\bar
p_Y$ be the uniform distribution over $S_Y\backslash\{\mathbf{y}_0\}$.
Similar to the reasoning of \eqref{eq:10}, we have
\begin{equation*}
    \sum_{\bx,\by} p_X(\bx)\bar p_Y(\by)\bx^{\transpose}\by \leq
  \sum_{\bx,\by}p_X(\bx)p_Y^*(\by)\bx^{\transpose}\by \leq nc.
\end{equation*}
Again, according to Lemma \ref{l3},
\begin{eqnarray*}
P_n^\dagger \leq P_n'& \leq &
\left(P_X \max_{\by}\bar p_Y(\by)\right)^{1/n} \\
& = & \left(P_X \frac{1}{\lfloor
      1/P_Y\rfloor}\right)^{1/n} \\
& \leq & \left(P_X \frac{1}{\lceil
      1/P_Y\rceil -1 }\right)^{1/n} \\
& \leq & (3P_XP_Y)^{1/n} \\
&=&\sqrt[n]{3}P_n^\dagger,
\end{eqnarray*}
and hence $\lim_{n\rightarrow\infty} P_n'/P_n^\dagger = 1$.

\begin{lemma}\label{l3}
For every $x\in(0,1)$,
\begin{equation*}
x(\lceil 1/x \rceil-1)\geq \frac{1}{3}.
\end{equation*}
\end{lemma}
\begin{IEEEproof}
If $x \geq \frac{1}{3}$, then
\begin{equation*}
x(\lceil 1/x \rceil-1)\geq x \geq \frac{1}{3}.
\end{equation*}
If $x<\frac{1}{3}$, then
\begin{equation*}
x(\lceil 1/x \rceil-1)\geq x(1/x-2)\geq 1-2x>\frac{1}{3}.
\end{equation*}
\end{IEEEproof}

\subsection{Proof of Theorem \ref{t3}}

We first show that we only need to consider $S_X$ and $S_Y$ with
profiles $a,b\in [0,0.5]^n$.  Suppose that for some \(i\) we have
\(a_i>\frac{1}{2}\).  We obtain a new set $S_X'$ by
flipping the $i$-th bit of all vectors in $S_X$.  Let $a' =
\Gamma(S_X')$. We have $a_k'=a_k$ for $k\neq i$ and $a_i'=1-a_i$.  We
know from Lemma~\ref{lemma4} that for the constraint
\eqref{constraint2} still holds with $S_X'$ in place of $S_X$ since
$a_i'<0.5< a_i$.  While the objective function of
Problem~\ref{problem2} with $S_X'$ in place of $S_X$ does not change
since $|S_X'|=|S_X|$. Similarly we can modify $S_Y$ such that all
\(b_i\leq \frac{1}{2}\).

Without the loss of generality, we assume $a_1\geq a_2\geq\cdots\geq
a_n$. Otherwise we just change the order of the bit in the string.
Now we put $b_1,...,b_n$ in a non-decreasing reordering as:
$b_1'\leq...\leq b_n'$. There must exist set $S_Y'\subseteq \{0,1\}^n$
such that $ \avg(S_Y')=(b_1',...,b_n')^{\transpose}$ by changing the
order of the bits for each string in set $S_Y$. Then we have
\begin{equation}
\frac{1}{n|S_X||S_Y'|}\sum_{x\in S_X,y\in S_Y'}x^{\transpose}y=\sum_{i=1}^{n}{a_ib_i'}\leq \sum_{i=1}^{n}{a_ib_i}\leq c.
\end{equation}
The proof is completed by $|S_X||S_Y'|=|S_X||S_Y|$.

\subsection{Proof of Theorem \ref{theorem1}}

The logarithm in this proof has base $2$.
Consider subset $S\subset \{0,1\}^n$ with $\avg(S)\leq a$. Define a
random vector $X=(X_1,X_2,\ldots,X_n)$ over $\{0,1\}^n$ with support $S$ and
$\Pr\{X=\bx\} = \frac{1}{|S|}$ for each $\bx\in S$.
Recall that the $i$-th component of $\bx \in \{0,1\}^n$ is denoted by $\bx_i$.
Let $l_k=\lfloor\frac{kn}{m}\rfloor$ for $k=0,1,\ldots,m$.
Since $(\text{E}[X_1],\text{E}[X_i],\ldots,\text{E}[X_n]) = \avg(S) \leq a$, we have for $k=1,\ldots,m$ and $i=1,\ldots,l_k-l_{k-1}$,
$\text{E}[X_{l_{k-1}+i}]=f_{\Gamma(S)}(\frac{l_{k-1}+i}{n}) \leq f_{a}(\frac{l_{k-1}+i}{n})=a_k$. Note that $X_i$ is a
binary random variable. Hence the entropy $H(X_{l_{k-1}+i}) \leq
\hb(a_k)$ for $k=1,\ldots,m$ and $i=1,\ldots,l_k-l_{k-1}$.
Therefore,
\begin{eqnarray*}
  \log |S| = H(X) & \leq & \sum_{k=1}^m \sum_{i=1}^{l_k-l_{k-1}}H(X_{l_{k-1}+i}) \\
  & \leq & \sum_{k=1}^m (l_k-l_{k-1}) \hb(a_k)  \\
  & \leq & \frac{n}{m} \left(\sum_i \hb(a_i)+o(1)\right),
\end{eqnarray*}
where the
last inequality follows from $l_k-l_{k-1} \leq \frac{n}{m}+1$ and
$o(1)$ tends to zero as $n$ tends to $\infty$.
Since the above inequality holds for all subset $S\subset \{0,1\}^n$
with $\avg(S)\leq a$, we have
\begin{equation*}
  V_n(a) \leq 2^{\frac{n}{m} (\sum_i \hb(a_i)+o(1))}.
\end{equation*}

\subsection{Proof of Theorem \ref{maintheorem}}

Let \(a=\avg(S_X)\), \(b=\avg(S_Y)\). By Theorem \ref{t3}, it is
sufficient for us to consider $S_X$ and $S_Y$ such that $0.5\geq a_1
\geq \ldots \geq a_n \geq 0$ and $0\leq b_1 \leq \ldots \leq b_n \leq
0.5$. Hence \(f_a\) is decreasing on \([0,1]\), and \(f_b\) is increasing
on \([0,1]\).

Define two \textit{m-profile}s \(\bar{a}\) and \(\underline{a}\) such
that for \(1\leq i\leq m\),
\begin{equation*}
  \bar a_i=\frac{\lceil mf_a\left(\frac{i-1}{m}\right)
\rceil}{m}, \
  \underline a_i= \frac{\lfloor mf_a\left(\frac{i}{m}\right)
\rfloor}{m}.
\end{equation*}
We have \(f_{\bar a}\) and \(f_{\underline{a}}\) are decreasing on
\([0,1]\).

\begin{lemma}\label{p1}
\(\underline{a}\leq a\leq \bar{a}\).
\end{lemma}

\begin{IEEEproof}
  Notice that \(f_a\) is a decreasing function.  For every \(0\leq
  r\leq 1\),
\begin{equation*}
f_{\bar{a}}(r)=\bar{a}_{\lceil rm\rceil}\geq f_a\left(\frac{\lceil rm\rceil-1}{m}\right)\geq f_a(r),
\end{equation*}
and similarly,
\begin{equation*}
f_{\underline{a}}(r)=\underline{a}_{\lceil rm\rceil}\leq f_a\left(\frac{\lceil rm\rceil}{m}\right)\leq f_a(r).
\end{equation*}
Thus \(\underline{a}\leq a\leq \bar a\).
\end{IEEEproof}

Define two \textit{m-profile}s \(\bar{b}\) and \(\underline{b}\) such
that for \(1\leq i\leq m\),
\begin{equation*}
  \bar b_i=\frac{\lceil mf_b\left(\frac{i}{m}\right)
\rceil}{m}, \
  \underline b_i= \frac{\lfloor mf_b\left(\frac{i-1}{m}\right)
\rfloor}{m}.
\end{equation*}
We have \(f_{\bar{b}}\) and \(f_{\underline{b}}\) are increasing on \([0,1]\),
and similar to Lemma~\ref{p1}, we have the following lemma.

\begin{lemma}
\(\underline{b}\leq b\leq \bar{b}\).
\end{lemma}

Now we can prove the following lemma.

\begin{lemma}\label{l10}
For $m\geq 2$,
\begin{equation}
\frac{1}{m}\sum_{i=1}^{m}{\bar{a}_i\bar{b}_i}-\frac{1}{n}\sum_{i=1}^{n}a_ib_i<\frac{2}{m}.
\end{equation}

\end{lemma}

\begin{IEEEproof}
Observe that
\begin{IEEEeqnarray*}{rCl}
  \frac{1}{m}\sum_{i=1}^{m}{\bar{a}_i\bar{b}_i}-\frac{1}{n}\sum_{i=1}^{n}a_ib_i&=&
  \frac{1}{m}\sum_{i=1}^{m}{\bar{a}_i\bar{b}_i}-\int_{0}^{1}{f_a(t)f_b(t)}\mathrm{d}t\\
  &\leq & \frac{1}{m}\sum_{i=1}^{m}{\bar{a}_i\bar{b}_i} -
  \int_{0}^{1}{f_{\underline{a}}(t)f_{\underline{b}}(t)}\mathrm{d}t \\
  &=&\frac{1}{m}\sum_{i=1}^{m}{\bar{a}_i\bar{b}_i}-\frac{1}{m}\sum_{i=1}^{m}{\underline{a}_i\underline{b}_i},
\end{IEEEeqnarray*}
where in the first and last equality we apply Lemma~\ref{lm:1}.
The first equality comes from the fact that \(f_a\) and \(f_b\) are step functions. By definition, we have for $1\leq i\leq m-1$, $m\underline{a}_{i}\geq
m\bar{a}_{i+1}-1$ and $m\underline{b}_{i+1} \geq m\bar{b}_{i}-1$. Hence
\begin{IEEEeqnarray*}{rCl}  
\IEEEeqnarraymulticol{3}{l}{\frac{1}{m}\sum_{i=1}^{m}{\bar{a}_i\bar{b}_i}-\frac{1}{m}\sum_{i=1}^{m}{\underline{a}_i\underline{b}_i}}
\\
  & \leq &
  \frac{1}{m}\sum_{i=1}^{m}\bar{a}_i\bar{b}_i-\frac{1}{m}\sum_{i=2}^{m-1} \left(\bar{a}_{i+1}-\frac{1}{m}\right)\left(\bar{b}_{i-1}-\frac{1}{m}\right)\\
  & = &
  \frac{1}{m}\Bigg(\bar{a}_1\bar{b}_1+\bar{a}_2\bar{b}_2+\sum_{i=3}^{m}\bar{a}_i(\bar{b}_i-\bar{b}_{i-2})+
 \\
& & \quad \quad  \sum_{i=2}^{m-1}\left(\frac{\bar{a}_{i+1}}{m}+\frac{\bar{b}_{i-1}}{m}-\frac{1}{m^2}\right)\Bigg)
  \\
  & \leq &
  \frac{1}{m}\Bigg(0.25+0.25+\sum_{i=3}^{m}0.5(\bar{b}_i-\bar{b}_{i-2})+
  \\
  & & \quad \quad \sum_{i=2}^{m-1}\left(\frac{0.5}{m}+\frac{0.5}{m}\right)\Bigg)
  \\
  & = &
  \frac{1}{m}\left(1.5-\frac{2}{m}+0.5\bar{b}_m+0.5\bar{b}_{m-1}-0.5\bar{b}_2-0.5\bar{b}_{1}\right) \\
  & \leq & \frac{2}{m},
\end{IEEEeqnarray*}
where we use the fact that $\bar{a}_i,\bar{b}_i \leq 0.5$.
\end{IEEEproof}

By Lemma \ref{l10} and the condition of the theorem (using the form
given in Lemma~\ref{lemma4}),  we have
\begin{equation}\label{eq:12}
\frac{1}{m}\sum_{i=1}^{m}{\bar{a}_i\bar{b}_i}\leq \frac{1}{n}\sum_{i=1}^{n}{a_ib_i}+\frac{2}{m}\leq c_Q+\frac{2}{m}.
\end{equation}
From  Lemma \ref{volumelemma} and Theorem \ref{theorem1}, we know that
\begin{IEEEeqnarray*}{rCl}
|S_X||S_Y| & = & V_n(a)V_n(b) \\
& \leq & V_n(\bar{a})V_n(\bar{b}) \\
& \leq & 2^{\frac{n}{m}(\sum_{i=1}^{m}{\left(\hb(\bar{a}_i)+\hb(\bar{b}_i)\right)}+o(1))},
\end{IEEEeqnarray*}
where $o(1)\rightarrow 0$ as $n\rightarrow \infty$.
For $0\leq t \leq 0.25$, define
\begin{equation}
  \label{eq:11}
  f(t)=\max_{2t\leq x\leq \frac{1}{2}}{\left(\hb(x)+\hb\left(\frac{t}{x}\right)\right)}.
\end{equation}
Some properties of the above function are given in
Appendix~\ref{sec:convexity-function} (see Lemma \ref{lemma7} -- \ref{concavelemma}).
We have
\begin{IEEEeqnarray*}{rCl}
\frac{1}{m}\sum_{i=1}^{m}\left(\hb(\bar{a}_i)+\hb(\bar{b}_i)\right)
& = &
\frac{1}{m}\sum_{i=1}^{m}\left(\hb(\bar{a}_i)+\hb\left(\frac{\bar{a}_i\bar{b}_i}{\bar{b}_i}\right)\right) \\
& \leq & \frac{1}{m}\sum_{i=1}^{m}f(\bar{a}_i\bar{b}_i)\leq f\left(c_Q+\frac{2}{m}\right),
\end{IEEEeqnarray*}
where the first inequality follows from the definition of
$f(\bar{a}_i\bar{b}_i)$ and the second inequality is obtained by
applying \eqref{eq:12} and Lemma \ref{concavelemma}.

Thus for any sufficiently large $m$,
\begin{equation*}
\liminf_{n\to\infty}{\frac{1}{\sqrt[n]{|S_X||S_Y|}}}\geq 2^{-f\left(c_Q+\frac{2}{m}\right)}.
\end{equation*}
Take \(m\to\infty\) we have
\begin{equation}
\liminf_{n\to\infty}{\frac{1}{\sqrt[n]{|S_X||S_Y|}}}\geq 2^{-f(c_Q)}=4^{-\hb(\sqrt{c_Q})},
\end{equation}
where the last equality is implied by Lemma \ref{lemma7}.

\subsection{Proof of Theorem \ref{finaltheorem}}

For every $n$, let 
\begin{equation*}
  S_X=S_Y=\{\bx\in\{0,1\}^{n}:\bx\textrm{ includes
  at most }n\sqrt{c_Q}\ 1s\}.
\end{equation*}
Then
\begin{equation}
|S_X|=|S_Y|=\sum_{i=0}^{\lfloor n\sqrt{c_Q}\rfloor}{\binom{n}{i}}=2^{n(\hb(\sqrt{c_Q})+o(1))},
\end{equation}
where $o(1)\rightarrow 0$ as $n\rightarrow\infty$.
Thus
\begin{equation*}
\lim_{n\to\infty}\frac{1}{\sqrt[n]{|S_X||S_Y|}}=\frac{1}{2^{2\hb(\sqrt{c_Q})}}=4^{-\hb(\sqrt{c_Q})}.
\end{equation*}
From the constructions of $S_X$ and $S_Y$, we know that
\begin{equation*}
\avg(S_X)=\avg(S_Y)\leq\left(\sqrt{c_Q}, \sqrt{c_Q},\cdots,\sqrt{c_Q}\right).
\end{equation*}
Therefore
\begin{IEEEeqnarray*}{rCl}
\frac{1}{n}\sum_{\bx\in S_X, \by\in S_Y}
\frac{1}{|S_X||S_Y|}\bx^{\transpose}\by & = & \frac{1}{n}
(\avg(S_X))^{\transpose}\avg(S_Y) \\
& \leq & \frac{1}{n}\sum_{i=1}^{n}{(\sqrt{c_Q})^2} \\
& = & c_Q.
\end{IEEEeqnarray*}
Thus \(S_X\) and \(S_Y\) satisfies constraints in Theorem \ref{t2}.

\section{Concluding Remarks}
\label{sec:concluding-remarks}

In this paper, we determine for
Problem~\ref{problem1} that when $c=c_Q$
\begin{equation}
  \label{eq:16}
  \lim_{n\rightarrow \infty} P_n = 4^{-\hb(\sqrt{c})},
\end{equation}
which is of particular interest for quantum
information. Note that our technique also shows that \eqref{eq:16}
holds for $c_Q \leq c < 1/4$. However, the existing technique in this
paper does not imply \eqref{eq:16} for $c < c_Q$, which holds if we
can show that $f(t)$ (defined in \eqref{eq:11}) is concave in
$[0,0.25]$. But we can only show the concavity of $f(t)$ for the range $[0.0625,
0.25]$ (see Appendix~\ref{sec:convexity-function}). Whether $f(t)$ is concave in
$[0,0.25]$ is of certain mathematical interest.

\section*{Acknowledgments}
  We thank Xiongfeng Ma, Xiao Yuan and Zhu Cao for introducing us this
  problem and providing insightful comments to our work. 

\appendices

\section{Background of the Optimization Problem}
\label{sec:deriv-optim-probl}

\subsection{CHSH Inequality}

A Bell test experiment has two spatially separated parties, Alice and
Bob, who can randomly choose their devices settings $X$ and $Y$ from
set $\{0,1\}$ and generate random output bits $A$ and $B$, respectively. The
Clauser-Horne-Shimony-Holt (CHSH) inequality is that
\begin{equation}\label{eq:1}
  S^{(1)} := \sum_{a,b,x,y\in\{0,1\}}(-1)^{a\oplus b + xy} q_{AB|XY}(a,b|x,y) \leq 2,
\end{equation}
where $\oplus$ denotes the exclusive-or of two bits, and $q_{AB|XY}(a,b|x,y)$
is the probability that outputs $a$ and $b$ are
generated when the device settings are $x$ and $y$. To simplify the
notations, we may also write $q_{AB|XY}(a,b|x,y)$ as $q(a,b|x,y)$, and
use the similar convention for other probability distributions. The theory of
quantum mechanics predicts a maximum value for $S$ of $S_Q=2\sqrt{2}$.

In a local hidden variable model (LHVM), assume that an adversary Eve
controls a variable $\lambda$ taking
discrete values so that
\begin{equation*}
  q(a,b|x,y) = \sum_{\lambda} q(a|x,\lambda)q(b|y,\lambda) q(\lambda|x,y),
\end{equation*}
where $q(a|x,\lambda)$ (resp.\ $q(b|y,\lambda)$) is the probability
that $a$ is output when the setting of Alice (resp. Bob) is $x$ (resp.
$y$), and $q(\lambda|x,y)$ is the conditional probability distribution
of the variable $\lambda$ given $x$ and $y$.  \emph{Free will} is
assumed in the derivation of the CHSH inequality, i.e.,
\begin{equation}\label{eq:14}
  q(\lambda|x,y) = q(\lambda).
\end{equation}
With this assumption, the inequality \eqref{eq:1} holds for any LHVM.

We consider the case that the device settings may not be chosen
freely, i.e., \eqref{eq:14} may not hold. By the Bayes' law,
\begin{equation*}
  q(\lambda|x,y) = \frac{q(x,y|\lambda) q(\lambda)}{q(x,y)}
  = 4 q(x,y|\lambda) q(\lambda),
\end{equation*}
where $q(x,y)$ is assumed to be $1/4$ so that Alice and Bob cannot
detect the existence of adversary Eve. In this case,
\begin{equation}\label{eq:4}
  S =   \sum_{\lambda} S_\lambda q(\lambda),
\end{equation}
where
\begin{equation*}
  S_{\lambda} = 4 \sum_{a,b,x,y\in\{0,1\}} (-1)^{a\oplus b + xy}
  q(a|x,\lambda) q(b|y,\lambda) q(x,y|\lambda).
\end{equation*}
The adversary can pick probabilities $q(\lambda)$, $q(x,y|\lambda)$,
$q(a|x,\lambda)$ and $q(b|y,\lambda)$ to fake the violation of a Bell's
inequality.

The following randomness measure are used in literature \cite{Koh12,Pop13,yuan2014randomness}
\begin{equation*}
  P = \max_{x,y,\lambda}q(x,y|\lambda).
\end{equation*}
Note that $P$ takes values from $1/4$ to $1$. When
$P=1/4$, all the device settings are uniformly picked independent of
$\lambda$. When $P=1$, for at least one value of $\lambda$, the device
settings are deterministic.

We are interested in the minimum value of $P$ such that $S\geq S_Q$
for certain LHVMs in the independent device setting scenario, i.e., $q(x,y|\lambda)
= q(x|\lambda)q(y|\lambda)$.   In other words, we want to solve the following
problem
\begin{equation}\label{eq:2}
  \begin{array}{cl}
   \displaystyle{\min} & \max_{x,y,\lambda}q(x,y|\lambda)\\
  \text{s.t.} & \sum_{\lambda} S_\lambda q(\lambda) \geq S_Q, \\
  & \sum_{\lambda} q(x,y|\lambda)q(\lambda) = \frac{1}{4}, \\
    & q(x,y|\lambda) = q(x|\lambda)q(y|\lambda),
  \end{array}
\end{equation}
where the minimization is over all
the possible (conditional) distributions $q(\lambda)$, 
$q(a|x,\lambda)$, $q(b|y,\lambda)$ and $q(x,y|\lambda)$ with $q(x,y|\lambda)=q(x|\lambda)q(y|\lambda)$.
Due to the convexity of the constraints with respect to
$q(a|x,\lambda)$ and $q(b|y,\lambda)$, we can consider only
deterministic distributions $q(a|x,\lambda)$ and $q(b|y,\lambda)$
without changing the optimal value of \eqref{eq:2}. Let $a = a(x,\lambda)$ and
$b = b(y,\lambda)$. Rewrite
\begin{equation}\label{eq:15}
  S_{\lambda} = 4 \sum_{x,y\in\{0,1\}} (-1)^{a(x,\lambda)\oplus b(y,\lambda) + xy}
  q(x,y|\lambda).
\end{equation}

In the above formulations, only a \emph{single run} of the test is
performed. It is more realistic to consider that the device settings
in different runs are correlated, which is referred to as the
\emph{multiple-run} scenario, where the device settings
$\bx=(x_1,\ldots,x_n)^\transpose$ and
$\by=(y_1,\ldots,y_n)^\transpose$ in $n$ runs of the tests follow a
joint distribution $q(\bx,\by|\lambda)$. Similar to the discussion of
the single-run scenario, for multiple runs, we have the CHSH
inequality $S^{(n)}=\sum_{\lambda}S^{(n)}_{\lambda}q(\lambda) \leq 2$ with
\begin{IEEEeqnarray*}{rCl}
  S^{(n)}_{\lambda} & = & \frac{4}{n} \sum_{\bx,\by\in \{0,1\}^n} q(\bx,\by|\lambda) \sum_{i=1}^n
  (-1)^{a(x_i,\lambda)\oplus b(y_i,\lambda) + x_iy_i} \\
  & = & 4 \sum_{\bx,\by\in \{0,1\}^n} q(\bx,\by|\lambda)
    \Big[\pi(0,0|\bx,\by) (-1)^{a(0,\lambda)\oplus b(0,\lambda)}
        \\
  & & +
    \pi(0,1|\bx,\by) (-1)^{a(0,\lambda)\oplus b(1,\lambda)} \\
  &  & + \pi(1,0|\bx,\by) (-1)^{a(1,\lambda)\oplus b(0,\lambda)}
       \\
  & & +
    \pi(1,1|\bx,\by) (-1)^{a(1,\lambda)\oplus b(1,\lambda)+1} \Big]
       \\
  & = & 4 \sum_{x,y\in \{0,1\}} (-1)^{a(x,\lambda)\oplus
    b(y,\lambda)+xy}\pi(x,y|\lambda)  \IEEEyesnumber \label{eq:6}
\end{IEEEeqnarray*}
where $\pi(x,y|\bx,\by)$ is the fraction of $(x,y)$ pairs among the pairs
$(x_k,y_k), k=1,\ldots,n$, and
\begin{equation*}
  \pi(x,y|\lambda) = \sum_{\bx,\by\in \{0,1\}^n}
q(\bx,\by|\lambda) \pi(x,y|\bx,\by).
\end{equation*}
Note that \eqref{eq:6} shares the same form as \eqref{eq:15}.

Define the measure of measurement dependence for multiple runs as
\begin{equation*}
  P^{(n)} = \left(\max_{\bx,\by,\lambda}q(\bx,\by|\lambda)\right)^{1/n}.
\end{equation*}
Under the independent device setting condition that $q(\bx,\by|\lambda)  =
q(\bx|\lambda) q(\by|\lambda)$, the problem of interest now becomes
\begin{equation}\label{eq:9}
  \begin{array}{cl}
   \displaystyle{\min} & \displaystyle{\left(\max_{\bx,\by,\lambda} q(\bx,\by|\lambda)\right)^{1/n}} \\
   \text{s.t.} & \displaystyle{\sum_{\lambda}S^{(n)}_\lambda q(\lambda) \geq
                 S_Q} \\
    & \sum_{\lambda} q(\bx,\by|\lambda) q(\lambda) = \frac{1}{4^n}, \\
    & q(\bx,\by|\lambda)  = q(\bx|\lambda)  q(\by|\lambda),
  \end{array}
\end{equation}
where $S^{(n)}_\lambda$ is defined in \eqref{eq:6}. Note that when $n=1$,
\eqref{eq:9} becomes \eqref{eq:2}.

\subsection{Simplification}

We use the case $n=1$ to illustrate how to simplify the above
optimization problem. 

First, we determine the choice of the output functions $a(x,\lambda)$
and $b(x,\lambda)$ using the approach in \cite{Pop13}.
For a give value of $\lambda$, there are totally $16$ different pairs
of the output functions $(a,b)$.  Table~\ref{tab:outputfunction} lists
the eight possible output functions with $a(0,\lambda)=0$. It is not
necessary to consider the other eight possible output functions with
$a(0,\lambda)=1$ since they give the same set of $S_\lambda$ as listed
in the last column in Table~\ref{tab:outputfunction}. 
Since the output functions with index $1,2,3,4$ are better than the
output functions with index $5,6,7,8$, respective, we use the former
four choices of the output functions. 

\begin{table*}
\centering
\caption{\label{tab:outputfunction} Output function assignment.}
\setlength{\tabcolsep}{10pt}
\begin{tabular}{cccccc}
  \hline\hline
   & $a(0,\lambda)$ & $a(1,\lambda)$ & $b(0,\lambda)$ & $b(1,\lambda)$
  & $S_\lambda/4$ \\
  \hline
  1 & 0 & 0 & 0 & 0 & $q(0,0|\lambda)+q(0,1|\lambda)+q(1,0|\lambda)-q(1,1|\lambda)$ \\
  2 & 0 & 0 & 0 & 1 & $q(0,0|\lambda)-q(0,1|\lambda)+q(1,0|\lambda)+q(1,1|\lambda)$ \\
  3 & 0 & 1 & 0 & 0 & $q(0,0|\lambda)+q(0,1|\lambda)-q(1,0|\lambda)+q(1,1|\lambda)$ \\
  4 & 0 & 1 & 1 & 0 & $-q(0,0|\lambda)+q(0,1|\lambda)+q(1,0|\lambda)+q(1,1|\lambda)$ \\
  5 & 0 & 0 & 1 & 0 & $-q(0,0|\lambda)+q(0,1|\lambda)-q(1,0|\lambda)-q(1,1|\lambda)$ \\
  6 & 0 & 0 & 1 & 1 & $-q(0,0|\lambda)-q(0,1|\lambda)-q(1,0|\lambda)+q(1,1|\lambda)$ \\
  7 & 0 & 1 & 0 & 1 & $q(0,0|\lambda)-q(0,1|\lambda)-q(1,0|\lambda)-q(1,1|\lambda)$ \\
  8 & 0 & 1 & 1 & 1 & $-q(0,0|\lambda)-q(0,1|\lambda)+q(1,0|\lambda)-q(1,1|\lambda)$ \\
  \hline\hline
\end{tabular}
\end{table*}

With the choices of the output functions as specified above, the constraint $\sum_{\lambda} q(x,y|\lambda)q(\lambda) =
\frac{1}{4}$ is redundant. To show this, we consider a LHVM (denoted by $L^*$) with a constant
$\lambda$, and output functions $a^*(x)=b^*(y)=0$. (Other choices of $a^*(x)$ and $b^*(y)$ can
be shown similarly.)  We use $q^*(x,y)$
to denote the device setting distribution related to this LHVM. Define
a new LHVM (denoted
by $L$) with $\lambda = 0, 1, 2, 3$ and $q(\lambda)=1/4$ as follows: The
output functions are assigned according to Table~\ref{tab:output}, and
the device setting distributions are assigned according to
Table~\ref{tab:setting}. It can be verified that 
\begin{IEEEeqnarray*}{rCl}
  P & = & \max_{x,y\in\{0,1\},\lambda=\{0,1,2,3\}} q(x,y|\lambda)  \\
  & = &
  \max_{x,y\in\{0,1\}} q^*(x,y),
\end{IEEEeqnarray*}
and 
\begin{IEEEeqnarray*}{rCl}
S & = & \sum_{\lambda=\{0,1,2,3\}} q(\lambda) 4 \sum_{x,y\in\{0,1\}} (-1)^{a(x,\lambda)\oplus b(y,\lambda) + xy}
  q(x,y|\lambda) \\
  & = & q^*(0,0) + q^*(0,1) + q^*(1,0) - q^*(1,1).
\end{IEEEeqnarray*}
Hence, if LHVM $L^*$ achieves the optimal value of \eqref{eq:2}, so
does LHVM $L$, which has $q(x,y) = 1/4$.

\begin{table}
\centering
\caption{\label{tab:output} Output function assignment.}
\setlength{\tabcolsep}{10pt}
\begin{tabular}{ccccc}
  \hline\hline
  $\lambda$ & $a(0,\lambda)$ & $a(1,\lambda)$ & $b(0,\lambda)$ & $b(1,\lambda)$ \\
  \hline
  0 & 0 & 0 & 0 & 0\\
  1 & 0 & 0 & 0 & 1\\
  2 & 0 & 1 & 0 & 0\\
  3 & 0 & 1 & 1 & 0 \\
  \hline\hline
\end{tabular}
\end{table}

\begin{table}
\centering
\caption{\label{tab:setting} Assignment of the device setting
  distributions.}
\setlength{\tabcolsep}{10pt}
\begin{tabular}{ccccc}
  \hline\hline
  $\lambda$ & $q(0,0|\lambda)$ & $q(0,1|\lambda)$ & $q(1,0|\lambda)$ & $q(1,1|\lambda)$ \\
  \hline
  0 & $q^*(0,0)$ & $q^*(0,1)$ & $q^*(1,0)$ & $q^*(1,1)$ \\
  1 & $q^*(1,0)$ & $q^*(1,1)$ & $q^*(0,0)$ & $q^*(0,1)$ \\
  2 & $q^*(0,1)$ & $q^*(0,0)$ & $q^*(1,1)$ & $q^*(1,0)$ \\
  3 & $q^*(1,1)$ & $q^*(1,0)$ & $q^*(0,1)$ & $q^*(0,0)$ \\
  \hline\hline
\end{tabular}
\end{table}

Further, for each of the four pairs
of output functions with index $1,2,3,4$ in Table~\ref{tab:outputfunction}, the corresponding
$S_{\lambda}$ involves only one summands with negative
coefficient. Since the four probability masses $q(0,0|\lambda)$,
$q(0,1|\lambda)$, $q(1,0|\lambda)$ and $q(1,1|\lambda)$ are  symmetry, these
four pairs of output functions achieve the same optimal value. Here we use
$a(x,\lambda)=b(y,\lambda)=0$ so that
\begin{equation*}
  \sum_{\lambda} S^{(1)}_\lambda q(\lambda) = 4 - 8 q_{XY}(1,1).
\end{equation*}
With these simplifications, the above minimization problem becomes
\begin{equation}\label{eq:42}
  \begin{array}{cl}
   \displaystyle{\min} & \max_{x,y,\lambda}q(x,y|\lambda)\\
  \text{s.t.} & q_{XY}(1,1) \leq \dfrac{4-S_Q}{8},\\
    & q(x,y|\lambda) = q(x|\lambda)q(y|\lambda).
  \end{array}
\end{equation}

For any $\lambda$ and $c\in [0,0.5]$, let $P(c)$ be the minimum value
of $\max_{x,y} q(x,y|\lambda)$ such that $q(1,1|\lambda) \leq c,
  q(x,y|\lambda) = q(x|\lambda)q(y|\lambda)$.
Note that $P(c)$ does not depend on the choices of $\lambda$, and
$P(c)$ is a non-increasing function of $c$. It clear that if we use
only a constant $\lambda$ in \eqref{eq:42}, the optimal value is
$P(\frac{4-S_Q}{8})$. Now we show that it is sufficient to consider a
constant $\lambda$. Suppose that $q^*(x,y|\lambda)$ achieves the
optimal value of \eqref{eq:42}. Let $c_{\lambda} =
q^*(1,1|\lambda)$. By the first constraint of \eqref{eq:42}, we have 
  $\sum_{\lambda} q^*(\lambda) c_{\lambda} = \frac{4-S_Q}{8}$, which
implies the existence of certain $\lambda^*$ such that $c_{\lambda^*} \leq
\frac{4-S_Q}{8}$. 
By the definition of $P(c)$, we have
\begin{equation*}
  \max_{x,y} q^*(x,y|\lambda) \geq P(c_{\lambda}),
\end{equation*}
which implies
\begin{equation*}
  \max_{\lambda,x,y} q^*(x,y|\lambda) \geq \max_{\lambda}
  P(c_{\lambda}) \geq P(c_{\lambda^*})  \geq P((4-S_Q)/8).
\end{equation*}
In other words, using a LHVM with $\lambda$ taking multiple values
cannot achieve smaller optimal value than $P(\frac{4-S_Q}{8})$.
Hence, it is sufficient to consider a constant $\lambda$, and 
\eqref{eq:42} becomes
\begin{equation*}%
  \begin{array}{cl}
   \displaystyle{\min} & \max_{x,y}q(x)q(y)\\
  \text{s.t.} & q_{X}(1)q_{Y}(1)\leq \dfrac{4-S_Q}{8}
  \end{array}
\end{equation*}

Similar to the reasoning of the
single-run case, we can use a deterministic
strategy \(\lambda\) with \(a(x,\lambda)=b(y,\lambda)=0\),
and simplify problem \eqref{eq:9} to
\begin{equation*}
  \begin{array}{cl}
   \displaystyle{\min} & \displaystyle{\left(\max_{\bx,\by}q(\bx,\by)\right)^{1/n}} \\
   \text{s.t.} & \displaystyle{\frac{1}{n}\sum_{\bx,\by\in
                 \{0,1\}^n}q(\bx,\by)\bx^\transpose\by\leq
                 \frac{4-S_Q}{8}},\\
    & q(\bx,\by) = q(\bx) q(\by),
  \end{array}
\end{equation*}
which is \eqref{eq:7}. %

\section{Properties of a Function}
\label{sec:convexity-function}

We study some properties of the function $f(t)$ defined in
\eqref{eq:11}. Recall that
\begin{equation*}
  f(t)=\max_{2t\leq x\leq
    \frac{1}{2}}{\left(\hb(x)+\hb\left(\frac{t}{x}\right)\right)},\quad 0\leq t \leq 0.25.
\end{equation*}
The next lemma implies that $f(t) = 2\hb(\sqrt{t})$
for \(0.0625\leq t\leq 0.25\).

\begin{lemma}\label{lemma7}
For \(0.0625\leq t\leq 0.25\), \(2t\leq x\leq 0.5\), we have
\begin{equation*}\label{eq:7.1}
\hb(x)+\hb\left(\frac{t}{x}\right)\leq 2\hb(\sqrt{t}),
\end{equation*}
where the equality holds for $x=\sqrt{t}$. That is
$f(t)=2\hb(\sqrt{t})$ for $t \in [0.0625, 0.25]$.
\end{lemma}
\begin{IEEEproof}
Fix \(t\). Let \( u(x)=\hb(x)+\hb\left(\frac{t}{x}\right)\).
Observe that \( u(x)=u\left(\frac{t}{x}\right)\).
Thus it suffices to show \(u(x)\leq 2\hb(\sqrt{t})\) for
\(2t\leq x\leq \sqrt{t}\). Taking derivative on \(u\) we have
\begin{equation*}
u'(x)=-\log x+\log (1-x)+\frac{t}{x^2}\log\left(\frac{t}{x}\right)-\frac{t}{x^2}\log\left(1-\frac{t}{x}\right)
\end{equation*}
Let \(v(x)=-x\log x+x\log(1-x)\), we have
\begin{equation}\label{eq:7.2}
xu'(x)=v(x)-v\left(\frac{t}{x}\right)
\end{equation}
From \( t\geq\frac{1}{16}\) we have
\begin{equation}\label{eq:7.3}
\frac{t}{x}\geq\frac{1}{2}-x\geq\frac{1}{4}.
\end{equation}
We may verify that \(v\) is decreasing on \([0.25,0.5]\).
If \(x\geq0.25\), then \(xu'(x)\geq0\) since \(\displaystyle x\leq\frac{t}{x}\).
Otherwise, we may verify \(v(x)\geq v(0.5-x)\) for \(x\leq 0.25\).
Then apply (\ref{eq:7.3}) to (\ref{eq:7.2}) we have
\begin{equation}\label{eq:7.4}
xu'(x)=v(x)-v\left(\frac{t}{x}\right)\geq v(x)-v\left(0.5-x\right)\geq 0
\end{equation}
Therefore \(u\) is an increasing function on \([2t,\sqrt{t}]\),
which implies \(u(x)\leq 2\hb(\sqrt{t})\).
\end{IEEEproof}

\begin{lemma}
  Function \(f(t)\) is increasing on \(\displaystyle \left[0,0.25\right]\).
\end{lemma}
\begin{IEEEproof}
To show that \(f\) is increasing, fix any \(0\leq t_1<t_2\leq0.25\).
We write \( f(t_1)=\hb(x_1)+\hb(y_1)\) where \(x_1\)
maximizes \( \hb(x)+\hb\left(\frac{t_1}{x}\right)\)
for \(x\in[2t_1,0.5]\) and \(x_1y_1=t_1\). We know that
\(0\leq x_1,y_1\leq 0.5\). Find \(x_2\) and \(y_2\) such that
\( x_1\leq x_2\leq \frac{1}{2}\),
\( y_1\leq y_2\leq\frac{1}{2}\)
such that \(x_2y_2=t_2\). Therefore
\begin{equation*}
f(t_1)=\hb(x_1)+\hb(y_1)\leq \hb(x_2)+\hb(y_2)\leq f(t_2).
\end{equation*}
\end{IEEEproof}

\begin{lemma}\label{concavelemma}
For any \(c'\geq c_Q = {2-\sqrt{2}\over 4} \approx 0.1464 \),
if \(k\) real numbers \(\displaystyle t_1,t_2,\cdots,t_k\in\left[0,0.25\right]\)
such that \(\frac{1}{k}\sum_{i=1}^{k}{t_i}\leq c'\), we have
\begin{equation*}
\frac{1}{k}\sum_{i=1}^{k}{f(t_i)}\leq f(c').
\end{equation*}

\end{lemma}

\begin{IEEEproof}
Let \(f_0(t)=2\hb\left(\sqrt{t}\right)\), \(\displaystyle 0\leq t\leq
0.25\). From Lemma \ref{lemma7} \(\displaystyle f(t)=f_0(t)\) for
\(t\geq 0.0625\). Let \(f_1\) be the tangent line of \(f_0\) on
\(\left(0.14,f_0(0.14)\right)\). Notice that \(\hb(x)\) and
\(\sqrt{x}\) are both concave on their domains. We see that \(f_0(t)\)
is also concave on \([0,0.25]\). Observe that \(f_0\) is concave
and increasing on \(\left[0,\frac{1}{4}\right]\), we have \(f_1\)
is an increasing function, while for every \(\displaystyle t\in[0,0.25]\),
\(f_0(t)\leq f_1(t)\).

Let \(g(t)\) be a function defined on \(\left[0,0.25\right]\) such that
\begin{equation*}
g(t)=\begin{cases}
f_1(t)& 0\leq t\leq 0.14;\\
f_0(t)& 0.14<t\leq0.25.
\end{cases}
\end{equation*}
Observe that \(g\) is linear on \([0,0.14]\) and concave on \([0.14,0.25]\),
thus \(g\) is concave on \([0,0.25]\). For \(0\leq t<0.0625\),
\begin{IEEEeqnarray*}{rCl}
f(t) & \leq & f(0.0625) \\
 & = & f_0(0.0625) \ (= 1.623) \\
 & < & g(0) \ (= 1.630) \\
 & \leq & g(t).
\end{IEEEeqnarray*}
For \(0.0625\leq t\leq0.25\),
\begin{equation*}
f(t)=f_0(t)\leq g(t).
\end{equation*}
Thus \(g\) is always not smaller than \(f\). Take \(t_1',t_2',\cdots,t_k'\leq 0.25\)
such that \(t_i\leq t_i'\) for all \(1\leq i\leq k\), while
\(\frac{1}{k}\sum_{i=1}^{k}{t_i'}=c'\).
Applying Jensen's inequality we have
\begin{IEEEeqnarray*}{rCl}
\frac{1}{k}\sum_{i=1}^{k}{f(t_i)} & \leq &
\frac{1}{k}\sum_{i=1}^{k}{f(t_i')} \\
& \leq & \frac{1}{k}\sum_{i=1}^{k}{g(t_i')} \\
& \leq & 
g\left(\frac{1}{k}\sum_{i=1}^{k}{t_i'}\right) \\ 
& = & g(c') \\
& = & f(c'), 
\end{IEEEeqnarray*}
where the first inequality holds since \(f\) is increasing, the second
inequality holds since \(g\) is always no less than \(f\), and the
last equality follows from $c'\geq c_Q > 0.14$.
\end{IEEEproof}

\end{document}